# FAN AFFINITY LAWS FROM A COLLISION MODEL

**Shayak Bhattacharjee**

Department of Physics,
Indian Institute of Technology Kanpur,
Uttar Pradesh – 208016,
India.

\* \* \* \* \*

## ABSTRACT

The performance of a fan is usually estimated from hydrodynamical considerations. The calculations are long and involved and the results are expressed in terms of three affinity laws. In this work we use kinetic theory to attack this problem. A hard sphere collision model is used, and subsequently a correction to account for the flow behaviour of air is incorporated. Our calculations prove the affinity laws and provide numerical estimates of the air delivery, thrust and drag on a rotating fan.

\* \* \* \* \*

# INTRODUCTION

The fan is the commonest of all aerodynamic devices, being employed in as diverse situations as living rooms, factories and jet engines. Using fluid dynamics [1] to estimate the performance of a fan is surprisingly tricky however. The performance of a fan is standardly expressed in terms of three affinity laws stated as under :

AL-I The air delivery rate is proportional to its rotation speed and varies as the cube of its diameter.

AL-II The total or static pressure is proportional to the square of the diameter, the square of the rotation speed and the density of the air.

AL-III The total power is proportional to the cube of the rotation speed, the fifth power of the diameter, and is proportional to the density of air.

These laws follow from the fluid mechanical actuator disk model [2], [3] proposed by Rankine and Froude. In its simplest form, it carries out a one-dimensional flow analysis assuming the air to be incompressible and inviscid. Bernoulli's principle [4], conservation of mass and conservation of momentum are employed to obtain the fan performance in terms of the flow velocity at the disk. More complex formulations with rotating stream tube assumption can be done for a refinement of the basic results.

In this work we attempt to derive simple proofs of the fan affinity laws starting from the basics of classical mechanics. Fluid mechanics is completely eschewed in our derivation. As such, our treatment is easily accessible to the junior undergraduate and also to the more enterprising high school student. We also aim to calculate a numerical estimate of the performance of a fan given its geometrical parameters. The form of the affinity laws gives rise to the possibility of a derivation from dimensional analysis [5]-[6], but we show why such an approach is inconclusive. The relevant parameters can be readily estimated from common sense. They are as follows.

| Parameter | Symbol |
|---|---|
| Fan radius (half of the sweep) | $R$ |
| Blade chord length | $L$ |
| Number of blades | $n$ |
| Rotation speed | $\omega$ |
| Blade pitch angle | $\alpha$ |
| Density of air | $\rho$ |
| Viscosity of air | $\eta$ |

Fig. 1 below shows these parameters for a fan.

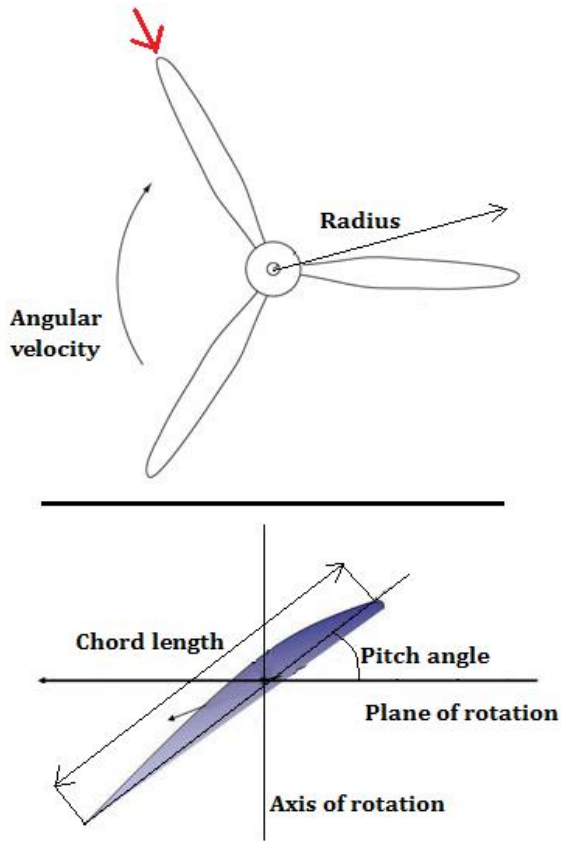

Figure 1 : *The relevant fan parameters used in the modelling. Upper panel shows front view of fan with the axis perpendicular to the plane of the paper. Lower panel shows the view along the arrow indicated in the upper panel. The cross section of the blade is visible. The plane of rotation extends into the paper (we see it in edge view) and hence the axis of rotation lies in the plane of the paper.*

Several problems are apparent. Firstly, the dimension of length can refer to either the chord length or the fan radius and it is not possible to differentiate between the two. Secondly the number of blades and the pitch angle are dimensionless so our analysis will not be able to provide the dependences on these quantities. Finally the realization that the entity $\rho\omega R^4/\eta$ is a dimensionless group makes dimensional analysis ineffective.

We now turn to kinetic theory. This theory models a gas as composed of innumerable small hard molecules which are moving around undergoing collisions either with each other or with the walls of the confining container. This model has led to successful predictions of material conductivity (Drude model) [7], viscosity and diffusion coefficient [8]. The exact numerical values obtained from such arguments are generally away from the actual values by factors of 2 or so but the order of magnitude as well as the dependences of these quantities on various parameters such as temperature, are predicted correctly. It is a reasonably standard exercise to calculate the viscous drag on a moving body from kinetic considerations – the problem set in [8] does it for a spacecraft and that of [9] for a sphere. We now extend these procedures to derive the fan affinity laws.

# DERIVATION

From a hard ball model viewpoint, the mechanism of fan action is through collisions of the air molecules with the fan blades. Let us compare the speed of the blades with that of the thermal motion of the air molecules. A typical fan rotates at say 1500 rpm and has a diameter of 60 cm. That gives a blade tip speed of about 50 m/s. On the other hand the thermal velocity of air molecule is given by $(3RT/M)^{1/2}$ where R is the universal gas constant, T is the Kelvin temperature and M is molar mass of air. This evaluates to about 450 m/s at a room temperature of 300 K. Hence we assume that the time scale of the collisions with the blades is much larger than that of the interparticle collisions. Because of this we can assume that at the time of collision, the particle velocities are random and average out to zero. In other words there is no drift velocity in the lab frame prior to the collision.

To get an estimate of the air delivery we want to find the volume of air sweeping through a surface parallel to the plane of the blades, per unit time. For this we need only the axial component of the particle velocity as it emerges from the blades, the tangential component plays no role here. We depict the collision in Fig. 2 below.

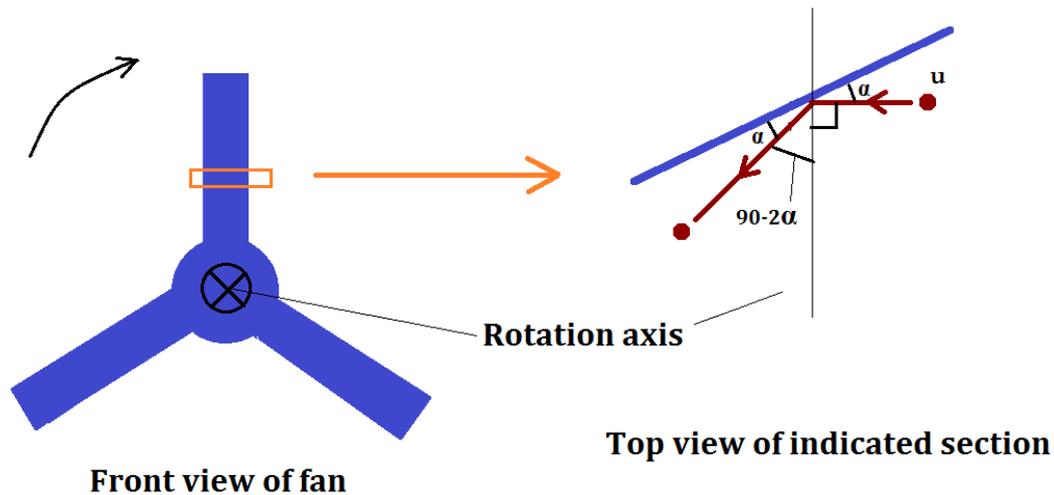

Figure 2 : *Front and top views of a fan blade, the latter showing a collision with an air molecule in the blade frame. The aerofoil section of Fig. 1 has been replaced by a simple section in the right panel here in accordance with our simple modeling.*

The left panel of Fig. 2 shows a rotating fan in the laboratory frame. The right panel zooms in on the section of blade marked in orange. This window is located at a radius of r and has a negligible extension in the radial direction. It covers the entire blade width though. In the lab frame this segment appears to move horizontally with speed

$$u = \omega r \qquad (1)$$

to the right. Hence, in a frame moving with the segment the air has a velocity of **u** towards the left, which has been shown in the figure. Since the blade is infinitely more massive than the air molecule, viewing the collision in the blade frame is easy; the air molecule simply follows the law of reflection like light off a mirror or a ball off a wall. This has been shown in the right panel of Fig. 2. The velocity component of interest is the one along the rotation axis. Easy geometry yields the angle values in the figure, and the axial component of the emergence velocity is seen to be **u sin2α**. This is in the blade frame. Since the motion of the blade is perpendicular to the axial direction, this component of the velocity will remain unchanged in the ground frame. Hence in the lab frame, the axial component of the exit velocity will also be **u sin2α**.

We now view the fan through a window at position **(r,θ)** and area **rdrdθ**. This window is shown in orange colour in Fig. 3.

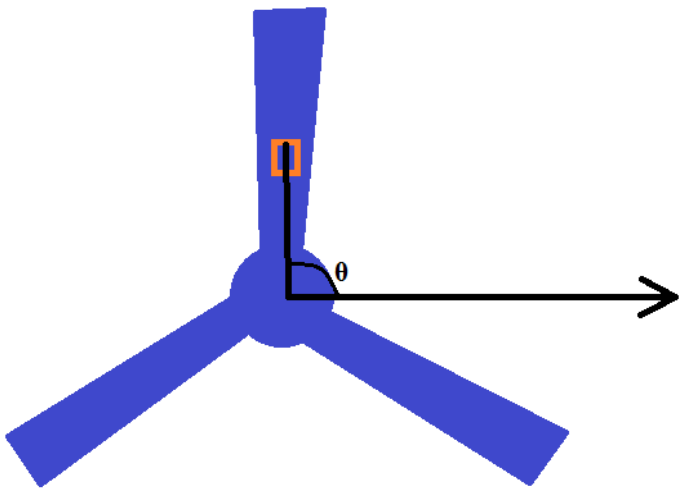

Figure 3 : *Viewing window.*

The volume rate **V** of flow through this window will be

$$dV = (\omega r \sin 2\alpha) r dr d\theta \quad . \tag{2}$$

We note that $\alpha$ might be a function of **r**. This in fact is the case in most real fans. As the mathematics gets messy if this feature is included, we simplify by assuming $\alpha$ to be constant. This constant is like the average blade pitch. The constant pitch assumption will be valid throughout the rest of our calculations. Eq. (2) must be integrated over all windows to obtain the final answer. At each instant of time, there will be airflow only from the region occupied by the blades and zero from the rest of the fan disc. This will set the limits on the integration over $\theta$. The angular span of a blade, i.e. the range of $\theta$ corresponding to each blade is approximately **(L cosα)/r**. Here the chord length **L** might be a function of **r**. Then the volume of displaced air is

$$V = n \int_0^R \int_0^{(L\cos\alpha)/r} \omega r^2 \sin 2\alpha \, d\theta \, dr \quad, \tag{3}$$

where n is the number of blades.

Application of this result to actual fans yields results which fall heavily short of the actual values. The fallacy occurs because it is assumed that the airflow does not occur in the region where there are no blades. This is contrary to experience. If air is suddenly set into motion, say by blowing, the motion persists for some time even after the cause is removed. This phenomenon has to be incorporated into our model. We also note that a single stimulus cannot cause air to flow forever; the flow velocity must die out in time. Combining these two phenomena we write the expression for airflow induced by a single stimulus (in this case the blade impact) at t=0 as

$$u(t) = (\omega r \sin 2\alpha) e^{-t/\tau} \quad . \tag{4}$$

The time constant can well be chosen to be large enough so that negligible decay can be assumed to occur in the time interval between successive passings of blades. This will allow the velocity to be taken the same at all $\theta$. The 'improved accuracy' from a more accurate modeling taking the damping into account will be nullified by the fact that the model itself is inaccurate. This renders futile the tedious calculations which result from the incorporation of the damping term. One might wonder why the damping was introduced in the first place – it is just for the sake of not proposing an unphysical concept like that of perpetual flow. Eq. (3) now becomes

$$V = \int_0^R \int_0^{2\pi} (\omega r^2 \sin 2\alpha) \, d\theta \, dr \tag{5}$$

which evaluates to

$$V = \frac{2\pi}{3} (\sin 2\alpha) \omega R^3 \quad . \tag{6}$$

This is nothing but the first affinity law. We apply Eq. (6) to a Crompton Greaves High Breeze industrial pedestal fan of diameter 450 mm, chord length approximately 6 cm, pitch angle about 18 degrees and a rotating speed of 1430 rpm. This yields the air delivery to be 124 m³ per minute, in excellent agreement with the rated value of 125 m³/min.

Encouraged by this result, we try to calculate the drag and thrust due to a rotating fan. This will be achieved by considering the change in momentum of the air after being struck by the blade. For this we need to calculate the mass of air being handled by the fan per unit time. We will again use rectangular viewing windows as in Fig. 3 but this time the window has to be oriented perpendicular to the flow so that the mass can be calculated correctly. This window is shown in Fig. 4.

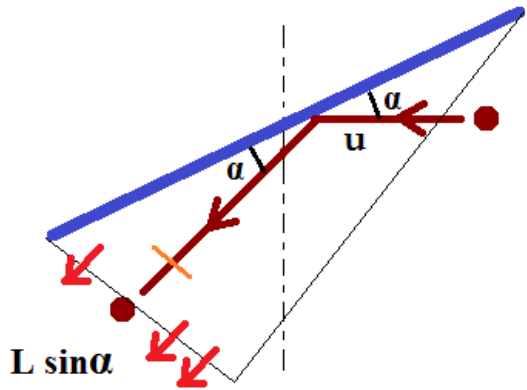

Figure 4: *Viewing window in orange. It extends vertically into the plane of the paper. The red arrows show the area through which airflow takes place. The width of this region is seen to be L sinα.*

One side of the window is along the radial direction, the second side is normal to the airflow. We will call this direction as **y**, whereby each window will cover an area **drdy**. Air flows through this window at a speed $\omega r$, hence the total flow volume through the window in a time **dt** is

$$d\upsilon = \omega r \, dr \, dy \, dt \qquad . \tag{7}$$

The mass flow in infinitesimal time is then

$$dm = \rho \omega r \, dr \, dy \, dt \qquad , \tag{8}$$

where we have introduced $\rho$, the density of air. For calculating the drag we are interested in the change in momentum in the direction of the blade motion, i.e. normal to the axial direction on account of the collision. This direction will hereafter be called the impingement direction. From the right panel in Fig. 2, the impingement component of the exit velocity in the blade frame is **u cos2α**. The transformation to the lab frame involves subtraction of the blade velocity **u**, hence the velocity component as seen from the lab frame is **u(-1+cos2α)**. Since the sign is arbitrary, we reverse it, taking care to be consistent later on. Thus the infinitesimal change in momentum

$$dp = dm\big(\omega r(1-\cos 2\alpha)\big) \qquad . \tag{9}$$

The force is dp/dt which evaluates to

$$dF = \rho\omega^2 r^2 (1-\cos 2\alpha) dr dy \tag{10}$$

and the torque, which is the product of the force and the radius, evaluates to

$$d\Gamma = \rho\omega^2 r^3 (1-\cos 2\alpha) dr dy \quad . \tag{11}$$

Now we decide on the limits of integration. The r limits are straightforward, running the length of the blade i.e. 0 to R. The y limits will be determined by the width of the region through which the flow takes place. Fig. 4 shows this to be 0 to L sin$\alpha$. Then we have, for n blades,

$$\Gamma = n \int_0^R \int_0^{L\sin\alpha} \rho\omega^2 (1-\cos 2\alpha) r^3 dy dr \quad . \tag{12}$$

For the simple but reasonably common case of L and $\alpha$ independent of r, this evaluates to

$$\Gamma = \frac{n}{4}\rho\omega^2 \sin\alpha (1-\cos 2\alpha) LR^4 \quad . \tag{13}$$

The power is the product of the angular velocity and the drag torque. We see that we have almost recovered the third affinity law except that one R of the law has been replaced in Eq. (13) by L. For the Crompton Greaves fan analysed above, this yields a drag torque of 0.15 Nm, which is considerably below the rated value of 0.6 Nm. Both these discrepancies are understandable as the flow property discussed after Eq. (3) has been ignored in this calculation. We now incorporate this property.

We assume that the additional change in momentum of this extra flow must also be due to the blades. Since all the airflow is being attributed to the blades, the effect will be the same as if there were blades all over the fan disc, each transferring momentum to the air. Such a continuum of blades is shown in Fig. 5.

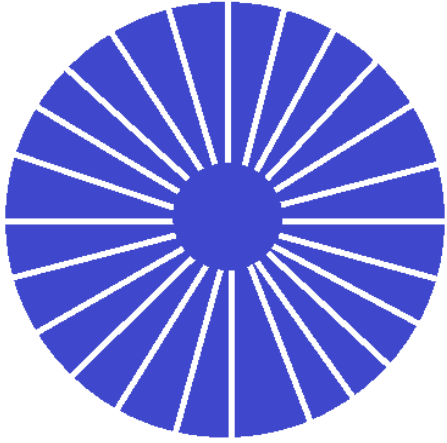

Figure 5 : *Hypothetical blade continuum to account for the residual airflow effect.*

We consider the shape of the blade to be such that it spans a small, constant angular displacement $\Delta\theta$. Then, geometrical considerations yield

$$L\cos\alpha = r(\Delta\theta) \tag{14}$$

and the r integral in Eq. (12) reduces to

$$\Gamma = \int_0^R n(\Delta\theta)\rho\omega^2(1-\cos 2\alpha)(\tan\alpha)r^4 dr \quad. \tag{15}$$

Now for the continuum of blades, n tends to infinity and $\Delta\theta$ tends to zero such that their product n($\Delta\theta$)=2$\pi$, whereby

$$\Gamma = \frac{2\pi}{5}\tan\alpha\left(1-\cos 2\alpha\right)\rho\omega^2 R^5 \quad. \tag{16}$$

This expression is identical to the third affinity law. For the fan treated previously, the result is 0.96 Nm which is 50 percent higher than the observed value. This level of inaccuracy is very common in kinetic theory and is an indicator of an accurate modelling.

Analogous to the drag, we can calculate the fan thrust. The calculation for the thrust will mirror that for drag except for the component of velocity under consideration. The drag features the impingement component in Eq. (9). The thrust must feature the axial component. As discussed after Eq. (1) this component is u sin2$\alpha$. Incorporating this change and carrying through the formalism leading to Eq. (16) from Eq. (9), the thrust evaluates to

$$T = \frac{4\pi}{5}\left(\sin^2\alpha\right)\rho\omega^2 R^5 \qquad . \tag{17}$$

The second affinity law obtains by recognizing that pressure is force per unit area. The total force is the sum of the drag and thrust components which have the same dependences on the various parameters, and the area is proportional to the square of the radius.

We note that the results derived by us are independent of the number of blades of the fan. This is a fallout of the crudity of the model. In reality the number of blades does affect the fan performance, which is why ceiling fans typically have 3 blades, exhaust fans 4 and modern propellers 6. A correction to account for this number can be included in our analysis by reconsidering our assumption of $\Delta\theta = 2\pi/n$ made before Eq. (16). A negative correction of order $1/n^2$ will arise from the fact that the integration treats arcs of circles as straight lines. However, we do not pursue this approach as other factors will play a stronger role in determining the n-dependence. The number of blades, and hence the air gap between successive blades, influences parameters like fan noise and energy dissipation during operation through vortex formation around the blades. Such an analysis cannot be carried out from our model, which, we have to admit, is not of much utility to a fan designer. The model should however be of considerable interest to the student of physics as it is a demonstration of how kinetic theory can almost trivially throw light on a very difficult problem.

## CONCLUSION

Thus we see that application of kinetic theory to the fan has produced accurate estimates of the air delivery, the thrust and the drag. Unlike the hydrodynamic derivation, our procedure is conceptually and mathematically simple. Nevertheless the three fan affinity laws have been proved and numerical estimates of fan performance have been obtained. The problem is thus a good demonstration of the strength and simplicity of kinetic theory.

## ACKNOWLEDGEMENT

I am grateful to KVPY, Government of India, for a generous Fellowship. I would also like to express my appreciation and gratitude to the anonymous reviewer for valuable comments which have substantially improved the quality of this paper. Finally, I would like to mention that the present problem arose while constructing a model for "Gearloose" event in Takneek 2011, the inter-Hall technical fest at IIT Kanpur.